\documentclass[fleqn,twoside]{article}
\usepackage[headings]{espcrc2}
\usepackage[ruled, vlined]{algorithm2e}
\usepackage{algorithmic}
\usepackage{epsfig}
\usepackage{balance}
\readRCS
$Id: espcrc2.tex,v 1.2 2004/02/24 11:22:11 spepping Exp $
\usepackage{graphicx}
\usepackage[figuresright]{rotating}

\newcommand{\AmS}{{\protect\the\textfont2
  A\kern-.1667em\lower.5ex\hbox{M}\kern-.125emS}}

\title{\textbf{Multiple Domain Secure Routing for Wireless Sensor Networks}}

\author{Lata B T, Jansi P K R, Shaila K, D N Sujatha, Venugopal K R\address[UVCE]{University Visvesvaraya College of Engineering, Bangalore, India, Contact: lata\_bt@yahoo.co.in }, and L M Patnaik\address[IISc]{Indian Institute of Science, Bangalore, India}}

\runtitle{ Multiple Domain Secure Routing for WSNs}
\runauthor{Lata B T et. al.,}

\begin{document}
\begin{abstract}
Secure Transmission of data packets in Wireless Sensor Networks is an important area of Research. There is a possibility of an attacker creating security holes in the network. Hence, network security and reliability can be achieved by discovering random multiple paths using multiple domains, and forwarding data packets from the source node to the destination node. We have designed, Multiple Domain Routing with Overlap of Nodes (MDRON) and Multiple Domain Routing Without Overlap of Nodes (MDRWON) algorithms, in which packets follow multiple optimized paths simultaneously. The Special node algorithm searches the node which has maximum power and these nodes are used for transferring the packet from one domain to another domain. Simulation results using MATLAB shows that performance is better than Purely Random Propagation (PRP) and Non Repetitive Random Propagation(NRRP) Algorithms.\\\\

\end{abstract}

\maketitle

\section{Introduction}
Wireless Sensor Networks (WSNs) consists of a large number of ultra-small, low-cost, battery-powered devices that have limited energy resources, computation, memory and communication capabilities. WSNs are often deployed in a vast terrain to detect events of interest and deliver data reports over wireless path to the sink. WSNs are used in environment monitoring, military applications, home appliance management, commercial applications, etc..\\\\
The major services of security are: Confidentiality, access control, authentication, non-repudiation and integrity. $Confidentiality$ : It means the data to be transmitted should be protected from passive attacks. \textit{Access Control} : It is the capability to control the access to host systems through the communication links. $Authentication$ : It assures that the communication is established between the authentic user. It provides confidence in the identity of the entities connected and assures that the received data is from the claimed source. $Nonrepudiation$ : It provides the proof that the message was sent or received from the specified device. $Integrity$ : It assures that the received data is exactly the same as it was sent by the authorized entity.\\\\
Even in the presence of security protocols, the adversaries are able to break the cipher text and find the plain text. Data security is essential for critical applications to work in hostile environments. It is easy for the attackers to inject malicious data messages or alter the content of legitimate messages during hop forwarding of data, due to the wireless environment in sensor networks. Thus, in sensor network communication, it is essential for authentication mechanisms to ensure that data from a valid source is not altered during transit. In this paper, we have focused on node compromise attack. \\\\
\textit{Motivation} : Data packets are usually forwarded from one node to another using single or multiple hops in WSNs. The adversary observes the traffic during routing and tries to compromise with the nodes, thus altering the data transfer, hence, the improvement over security and reliability has to be addressed.\\\\   
\textit{Contribution}: We have considered multiple domains for secure routing in WSNs. Special nodes will be discovered from the domains using special node generation algorithm. Nodes with maximum power availability in respective domains are considered as special nodes through which the domains communicate and secure communication link is established between any two domains. 
\subsection{Organization}
The rest of the paper is organized as follows: Section 2 discusses briefly the related work.The background of the work is explained in Section 3. Section 4 describes the architecture of the network model. Section 5 states the problem definition. The algorithms are discussed in Section 6. Section 7 gives the description of implementation while the performance of the algorithms is analysed in section 8. Conclusions are presented in Section 9.
\section{Literature Survey}
Yun et al., \cite{YY} discuss many security challenges, attacks, security requirements, key distribution schemes, authentication, integrity, secure routing, intrusion detection and countermeasures, active attacks, denial of service. Chris et al., \cite{CD} proposed threat models and security goals for secure routing in Wireless Sensor Networks. They introduce two novel classes of previously undocumented attacks against sensor networks-sinkhole attacks and Hello floods. They have shown, how attacks against ad-hoc wireless networks and peer-to peer networks are adapted into powerful attacks against sensor networks. Security analysis of all the major routing protocols and energy conserving topology maintenance algorithms are proposed for sensor networks. Chih-Yung et al., \cite{CCS} propose a protocol, (S-RGP) Active Route Guiding Protocol for a Single Obstacle that actively constructs the forbidden region for the concave region as soon as the obstacle is formed. The selected boarder nodes and forbidden-region boundary nodes prevent the packets from visiting the concave regions but guide the packet to the best route. The Route Guiding Protocol for Multiple (M-RGP) integrates the information of newly and previously formed obstacles and dynamically guides the packets to the best route.\\\\
Samundiswary et al., \cite{SSD} develops a Secure Greedy Perimeter Stateless Routing (S-GPSR) Protocol for Mobile Sensor Networks. In S-GPSR, Trust levels and neighbor hood distance are used for forwarding the packet based on the trusted distance, instead of minimal distance. Trust Update Interval (TUI) of each forwarded packet is buffered in the node. It determines a time, a node should wait before assigning a trust or distrust level to a node based upon the results of a particular event. After transmission, each node waits for the neighboring node to forward the packet. \\\\
Wenjing et al., \cite{WY} introduced H-SPREAD (A Hybrid Multipath Scheme for secure communication and Reliable Data Collection) is a hybrid multipath data collection scheme. It is based on distributed N-to-1 multipath discovery protocol. Multiple node-disjoint paths are discovered from every sensor node to the Base Station (BS) at the cost of some extra message exchanges. It is based on branch aware flooding and hence, provides security and reliability. It is resilient to node/link failures and a collusive attack of the compromised nodes. Feilong et al., \cite{FMMAM} propose two protocols MLR (Maximal Network Lifetime Routing) and SecMLR (Secure Maximal Network Lifetime Routing)  for Wireless Mesh Sensor Networks in pervasive environments.  MLR aims at maximizing network lifetime, merging the advantages of table driven routing protocols. In SecMLR Protocol, the sensor node sets up a routing table with multiple entries, and each of them routes data from the sensor node to a specified gateway.\\\\
\indent Abdul et al., \cite{AMC} focus on Wireless Sensor Network (WSNs) security and tolerance to the damage caused by an adversary which has compromised with deployed sensor node to modify, block or inject packets. Shu et al., \cite{SKL} introduced mechanisms considering single domain that generate randomized multipath routes. Routes taken by the parts of data of different packages change over time. Besides randomness, the generated routes are also highly dispersive and energy efficient, making them quite capable of overcoming black holes. Tian et al., \cite{TNJI} states a novel technique called $slicing$, which partitions the data both horizontally and vertically. This technique preserves better data utility. Another advantage of slicing is that it can handle high-dimensional data. Yihong et al., \cite{YJGMS} propose a Secure Sector Based Bi-path Clustering and Routing (SBBCR) protocol which controls energy consumption, provides higher level security for the network and effectively extend the network life time without performance degradation. \\\\
Anfeng et al., \cite{AZCZX} formulated the secret-sharing-based multipath routing problem, i.e., three-phase disjoint routing scheme called Security and Energy-efficient Disjoint Route (SEDR) as an optimization problem. Min-Ho et al., \cite{MYHS} proposed a new group key management scheme for multiple multicast groups, called the Master-Key-Encryption-based Multiple Group Key Management Scheme. It exploits asymmetric keys, $i.e.$, a master key and multiple slave keys, which are generated from the proposed master key encryption (MKE) algorithm, is used for efficient distribution of the group key. It reduces the storage and the communication overhead. Asmae et al., \cite{AANM} have discussed attacks such as eavesdropping, traffic analysis, disruption, hijacking, physical attacks such as an adversary may physically damage the hardware to terminate nodes, security goals, security challenges such as timing obfuscation, topology obfuscation, secure aggregation, aggregation with privacy.  
\section{Background}
Shu et al., \cite{SKL} developed four distributed schemes for propagating information, First, Purely Random Propagation (PRP), that uses only one-hop neighborhood information and provides baseline performance. Second, Directed Random Propagation (DRP), which utilizes two-hop neighborhood information to improve the propagation efficiency, leading to a smaller packet interception probability. Third, the Non Repetitive Random Propagation (NRRP), records all traversed nodes to avoid traversing them again in the future. Fourth, the Multicast Tree-assisted Random Propagation (MTRP), propagates the parts of data in the direction of the sink, making the delivery process more energy efficient.
\section{Problem Definition}
Consider a given Wireless Sensor Network consisting of $M$ sensor nodes. Let $c_{n}$ nodes compromise with the adversary in the same network due to node compromise attack, then, the packet delivery ratio decreases. Hence, security must be provided against compromised node attack and increase the packet delivery. The main objectives of the proposed work are: 
\begin{enumerate}
\item To randomly select special nodes to establish communication in multiple domain to avoid node compromise.
\item To discover multiple routes from source node to destination node, considering multidomain approach. 
\item To increase packet delivery ratio, reduce packet drop and increase throughput. 	
\end{enumerate}
\section{Network Model}
In this work,  Secure multipath routing algorithm is developed to overcome Compromised Node attacks. Instead of selecting paths from a pre-computed set of routes, the goal is to compute multiple paths in a randomized way each time an information packet needs to be sent. A large number of routes are generated for each source and destination. The goal is to compute multiple paths in a randomized way and establish a route between any two nodes present in two different domains.
\subsection{Assumptions}
\begin{enumerate}
\item Initially all nodes are randomly deployed.
\item Each node transmission range is 100-200 meters.
\item The node with maximum power availability are considered as special node. These nodes cannot be a compromised node.
\item Multihop relay is used if the intended destination is more than hop count away from the source. 
\end{enumerate} 
\section{Multiple Domain Routing Algorithms}
If the given network is considered to be a single domain then the node selects the neighbor node as the intermediate node or the node which had already traversed. This leads to looping of the path and the packets revolve within the loop or reach the destination node with a delay. Due to this, it is easy for the adversary to compromise with the looped node. In order to overcome this problem, the given region in a Wireless Sensor Networks are split into multiple domains. Each of the domain selects a special node randomly for every $n$ iterations. The nodes within one domain routes that data to another node in another domain using the special nodes. A multipath routing algorithm for multiple domains is developed that can overcome compromised-node attack. Instead of selecting paths from a pre-computed set of routes, the goal is to compute multiple paths in such a way, that each time it takes, a different path to transmit the information. To achieve this, an algorithm for Secure Multiple Domain concept is proposed for establishing the path between different domains through which the information is routed securely. The path is established with overlapping and without overlapping of nodes. In both the cases, the algorithm is implemented in two phases: ($A$) Special node selection Phase and ($B$) Path Selection(Tracing) Phase.
\begin{algorithm}
\KwIn{Battery power of all the nodes in each domain stored in two arrays A[ ], B[ ].}
\KwOut{Special Node is calculated having highest power SN=A[1];  //{\bf SN}:Special Node.}
//{\bf M} :maximum no of nodes in domain 1.\\
 //{\bf N} :maximum no of nodes in domain 2.
\BlankLine
{
\bf for {$i$ $=$ $2$ to $M$} do \\
{    
	\If{$A[i]$ $>$ $SN$}{
		$SN$=$A$[$i$]\;
		
	}
}

\Return $SN$\;
}
Repeat the above steps for domain 2, considering the array B[ ] instead of $A$[ ] and $N$ instead of $M$\;
\caption{Special Node Election}
\label{algo:specnode}
\end{algorithm}
\subsection{Special Node Selection Phase}
In the first phase, the $N$ nodes available in the region of interest is split into four domains. Each domain has uneven number of nodes which is the subset of $N$ nodes. Then, the special node is selected in each domain using game theory as shown in Table \ref{algo:specnode}. At a particular instant of time the power available in each node is determined. Then, the node with maximum power is selected as the special node as discussed in Algorithm \ref{algo:specnode}. 
\subsection{Path Selection Phases}
\subsubsection{Multiple Domain Routing with Overlap of Nodes (MDRON)}
In the second phase, the data to be transferred from the source node to the destination node in a particular coverage area is split into smaller packets by applying Linear Block Code (LBC) concept. A counter is used to keep track of the random selection of special node and special node table is updated regularly with the node IDs which were selected as special nodes. This table helps in avoiding the same node being selected again. The routing table is
\begin{algorithm}
{ {\bf variable}	:Source node($Sn$), Destination node($Dn$), Coverage area ($A$), Counter ($C$), Special Node ($Sp$), Intermediate Nodes ($in$)}\\
\KwIn{1. Routing Table of $Sn$ as $RT$[ ][ ] \\
                         2. Neighbors within the coverage area of the source node ($Sn$).\\
                         3. Neighbors routing table ($Ne$[ ][ ])\\
                         4. Source Neighbors list ($SN$[ ][ ])\\ }
\KwOut{Min$\_$hop$\_$route.}
\BlankLine
{\bf Label1:}\\
{\bf for} ($i$ $=$ $0$ to $n$) {\bf do}\\
// $n$ = Number of nodes in domain1 + number of nodes in domain2.\\
\If{($Dn$ $==$ $Sne[i]$)}{
	break; //stop because destination is reached.\;
      	\If{($sp$ $==$ $snc[i]$)}{
		Establish link between Domain1 and Domain 2.\;
    }
               c- -\;
               break\;
	       Pick the Node Randomly\;
    {
	  \If{(c != 0)}	{
             $Sn$ $=$ $in$ \; // in = intermediate nodes\\
  	     $Sne$[ ][ ] = $in$[ ][ ]\; // Intermediate neighboring list so becomes new source routing list.\\
	     goto $label1$\;
	}
}
            call LBC( ); // LBC-Linear Block Codes.\\
	    {Min$\_$hop$\_$Routing( )\;
}
}
end for\;
\Return Min$\_$hop$\_$route\;
\caption{MDRON}
\label{algo:MDRON}
\end{algorithm}
updated based on the neighboring nodes. It verifies if the neighboring node is a destination node, and the data packets are transferred to the destination.\\
If the node is other than the destination node, then, it checks if there exist a special node, then the data packets are transferred through these nodes from one domain to the other domain. If a special node does not exists then, it selects the neighbor node randomly, In both the cases, once the operation is performed, the counter value is decremented by one until it reaches zero. When the counter value is zero, the data packets are routed with minimum number of hops as discussed in Table \ref{table:outspec}. When the counter value is greater than zero, then, the source and destination nodes are assigned for a particular coverage area. In MDRON, the packets routing through the WSNs traverses through the same node thus, forming a loop sometimes. Due to this looping effect, there is a possibility of the data packets being lost or there might be a delay in the data packets reaching the destination, resulting in reduced packet delivery ratio. 
\subsubsection{Multiple Domain Routing Without Overlap of Nodes (MDRWON)}
The special node is selected in the first phase. In the second phase, the counter is initialized. The details of the neighboring nodes are updated in the routing table. If the selected node is a destination node, the data is transferred to the destination node. If the node is other than the destination node, it checks if there exists a special node. If it is not a special node, then the neighbor node is picked which is nearer to the special node. \\\\
The details of the selection of the node is compared with the routing list. If the node is earlier selected during the process of routing, then, its ID will be present in neighbor list field. It checks if any other node is available which is not used during the same trip and their ID is not available in the neighbor list field but should be present in neighbor list). If such a node is available, that new node acts as an intermediate node for transferring the data packets. At the same time, the routing table is updated and the value of counter is decremented. The present intermediate node becomes a source node and the process is repeated until the counter value reaches zero. When counter value reaches zero, minimum routing takes place. The algorithm for multiple domain routing without overlap of nodes is discussed in Algorithm \ref{algo:MDRWON}. In this case, it checks in the neighbor list if the packet has passed through the node, if it has taken this path earlier, then it takes a new route, thus eliminating the looping effect. 
\begin{algorithm}
{ {\bf variable}	:Source node($Sn$), Destination node($Dn$), Coverage area ($A$), Counter ($C$), Special Node ($Sp$), Intermediate Nodes ($in$)}\\
\KwIn{1. Routing Table of $Sn$ as $RT$[ ][ ]  \\
                 2. Neighbors within the coverage area of the source node ($Sn$).\\
                 3. Source Neighbors list ($SN$[ ][ ])\\
                 4. Neighbors routing table ($Ne$[ ][ ])\\ }
\KwOut{Min$\_$hop$\_$route.}
\BlankLine
{\bf Label1:}\\
{\bf for} ($i$ $=$ $0$ to $n$) {\bf do}\\
// $n$ = Number of nodes in domain1 + number of nodes in domain2.\\
\If{($Dn$ $==$ $Sne[i]$)}{
	break; //stop because destination is reached.\;
      	\If{($sp$ $==$ $snc[i]$)}{
		Establish link between Domain1 and Domain 2.\;
    
               c- -\;
               break\;
   }
   
	       select non-repetitive neighbor node by comparing the node in Route field ($RF$) of ($RT$) and $Ne$[ ][ ](neighbor list)\;
               c- -\;
    {
	  \If{(c != 0)}	{
             go to $label1$ \; 
    }{
  	     \If{($c$ $==$ $0$)}
		{call Min$\_$hop$\_$Routing( )\;
	}
}
}
}
end for\;
\Return Min$\_$hop$\_$route\;
\caption{MDRWON}
\label{algo:MDRWON}
\end{algorithm}
\section{IMPLEMENTATION}
The network consists of $N$ nodes, which are divided into four domains. The number of nodes in each domain is the subset of $N$. Each domain consists of uneven number of nodes. In all the four domains, each node broadcasts \textit{hello message} to update the routing table with its neighbor information within the domain. Every node sends the hello message to one  hop neighbor, before the first phase and there after the network is split into domains, and the path message is flooded to the neighboring nodes. \\\\
The neighbor node looks to the previous list and establishes the link between the source and the destination node. The location of the nodes within the domain is randomly changed for every $i$ iterations so that it is difficult for the adversary to compromise with any node. Again, the routing table is updated with its neighbor information and the path is established between the source and the destination nodes. When we discover a random path from source node to a destination node, it is very difficult for the attacker to know the path, since, the random path is unique. Even if the attacker is successful in compromising the node, next path selected is random for the next packet; hence, complete data cannot be explored by the attacker.
\section{Simulation and Performance Evaluation}
In this section, the secure path establishment in multiple domain is simulated using the MATLAB on Windows Platform.
\subsection{Simulation Setup}
The node topology consists of 100m x 100m with 250 sensors deployed randomly and the transmission range is set to 50$m$. Simulation is performed for 50 runs with 1500 sec of simulation time. 
\begin{table}
\label{table:outspec}
\caption{Output of Special Node Selection Algorithm} 
\centering
	\begin{tabular}{|c|c|c|} \hline
	{Iteration} & {Domain} & {Special}    \\ 
	{Number} & {Number} & {Node}    \\  \hline 
    	    & {1} &{29}	     \\ 	
	    & {2} &{69}	     \\ 
        {1} & {3} &{120}	     \\ 	
	    & {4} &{204}	     \\  \hline
	    & {1} &{10}	     \\ 	
	{2} & {2} &{70}	     \\ 
            & {3} &{102}	     \\ 	
	    & {4} &{304}	     \\  \hline
\end{tabular}
 \label{table:outspec}
\end{table}
\begin{figure}
\centering
\includegraphics[width=18pc, height=12pc]{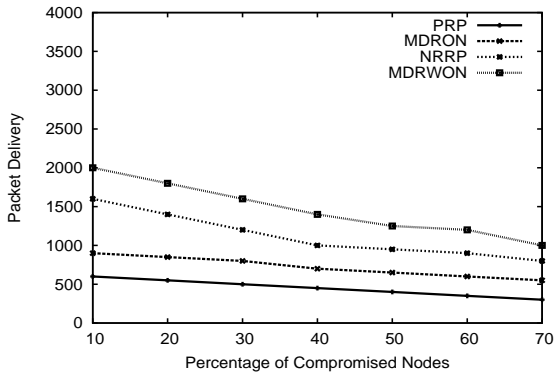}
\caption{Compromised Nodes Vs.Packet Delivery ratio.}
\end{figure}
\begin{figure}
\centering
\includegraphics[width=18pc, height=12pc]{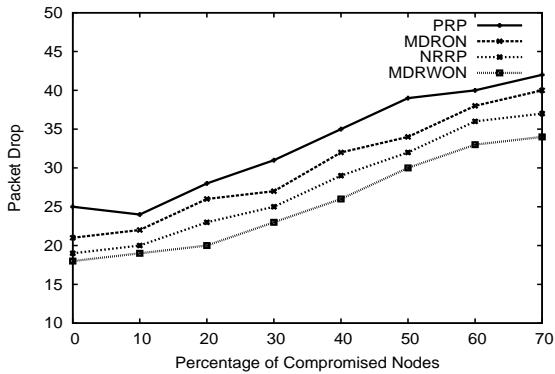}
\caption{Compromised Nodes Vs.Packet Drop.}
\end{figure}
\begin{figure}
\centering
\includegraphics[width=18pc, height=12pc]{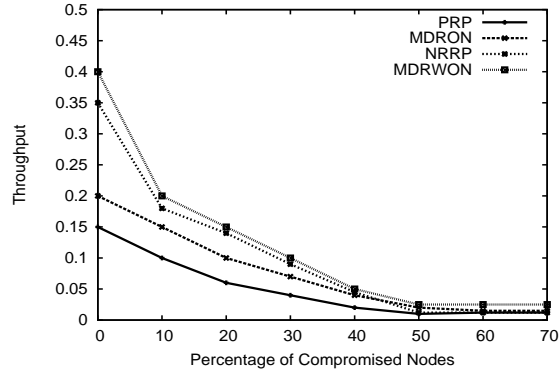}
\caption{Throughput Vs. Compromised Nodes}
\end{figure}
\subsection{Performance Evaluation}
In PRP and NRRP algorithms, routing is done without special nodes. In our proposed MDRON and MDRWON algorithms, special nodes are computed using special node algorithms as discussed in Algorithm 1. Random path includes special node in each domain, which in turn enhances the packet delivery ratio as compared with PRP and NRRP. Multiple domains are not considered in PRP and NRRP. We have proposed the algorithms for both single and multiple domains. The result for special node selection algorithm is available in Table \ref{table:outspec}.  Comparison of PRP, MDRON, NRRP and MDRWON is shown in Figure 1. The variation of packet drop for different number of compromised nodes in the network is shown in Figure 1. MDRON and MDRWON algorithms results better packet delivery ratio, as compared with PRP and NRRP. \\
\indent In Figure 2, when the percentage of the number of compromised nodes increase, the packet drop increases gradually. In MDRON, there is a chance of overlapping of nodes. In MDRWON, overlapping of nodes is avoided, hence, the packet drop reduces by 2\% in MDRWON as compared with MDRON. The packet delivery ratio decreases proportionately with the increase in the percentage of the number of compromised nodes as observed in Figure 2. The packet delivery ratio decreases by 8\% in MDRON compared with MDRWON. Correspondingly, the throughput of the network reduces gradually with increase in the percentage of compromised nodes as observed in Figure 3. The average delay of the packets reaching the destination increases with increase in the number of compromised nodes. The delay of the packets reaching the destination is reduced in MDRON and MDRWON by 10\% and 2\% compared with PRP and NRRP respectively as shown in Figure 4.
\begin{figure}
\centering
\includegraphics[width=18pc, height=12pc]{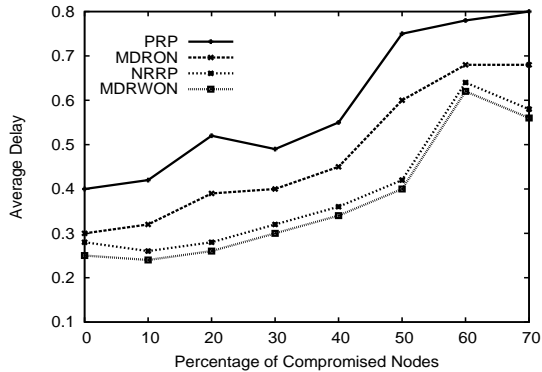}
\caption{Average Delay Vs. Compromised Nodes.}
\end{figure}
\section{CONCLUSIONS}
Security is an important issue in WSNs. This paper proposes a Secure Multiple Domain Routing for Wireless Sensor Networks, that consists of a special node, in each domain, which helps in establishing secure communication. The algorithm discovers the route from one domain to another domain and within the domain which minimizes the effect of node compromise. Here, We are communicating through special nodes, which inturn enhances the delivery ratio for multiple domains, as compared with previous algorithms, decreases packet drop that is caused by the compromised nodes, and increases throughput. Numerical results demonstrates that the performance is better interms of reduced packet drop, increases packet delivery and throughput by avoiding the node from compromising with the adversary.

\noindent {\includegraphics[width=1in,height=1.7in,clip,keepaspectratio]{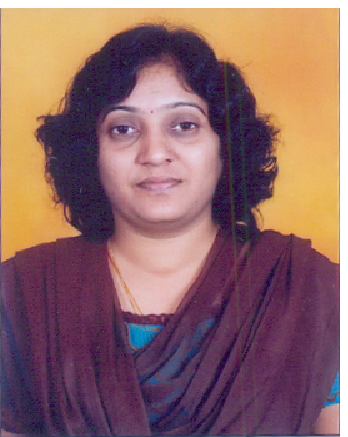}}
\begin{minipage}[b][1in][c]{1.8in}
{\centering{\bf {Lata B T}} is an Assistant Professor in the Department of Computer Science and Engineering at University Visvesvaraya College of Engineering, Bangalore Uni-versity, Bangalore, India. She obtained her B.E degree in }\\\\
\end{minipage}
Computer Science and Engineering from Karnatak University, Dharwad, and M.Tech. degree in Computer Network Engineering from Visveswaraiah Technological University, Belgaum. She is pursuing her Ph.D degree in the area of Wireless Sensor Networks in Bangalore University, Bangalore. Her research interest is in the area of Sensor Networks.\\\\

\noindent {\includegraphics[width=1in,height=1.7in,clip,keepaspectratio]{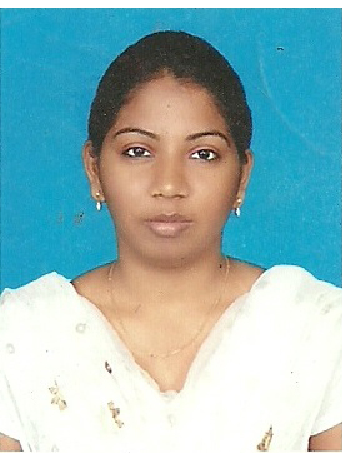}}
\begin{minipage}[b][1in][c]{1.8in}
{\centering{\bf {Jansi P K R}} obtained her M.E in the Department of Computer Science and Engineering at University Visvesvaraya College of Engineering, Bangalore Uni-versity, Bangalore, India. Her research interest is in the area of Sensor Networks.}\\
\end{minipage}\\\\
\noindent {\includegraphics[width=1in,height=1.7in,clip,keepaspectratio]{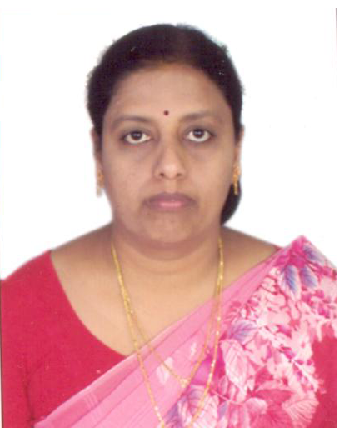}}
\begin{minipage}[b][1in][c]{1.8in}
{\centering{\bf {Shaila K}} is an Professor and Head in the Department of Electronics and Communication Engineering at Vivekananda Institute of Technology, Bangalore, India. She obtained her B.E in Ele-}\\
\end{minipage}
ctronics and M.E degrees in Electronics and Communication Engineering from Bangalore University, Bangalore. She obtained her Ph.D degree in the area of Wireless Sensor Networks in Bangalore University. Her research interest is in the area of Sensor Networks, Adhoc Networks and Image Processing.\\\\

\noindent {\includegraphics[width=1in,height=1.7in,clip,keepaspectratio]{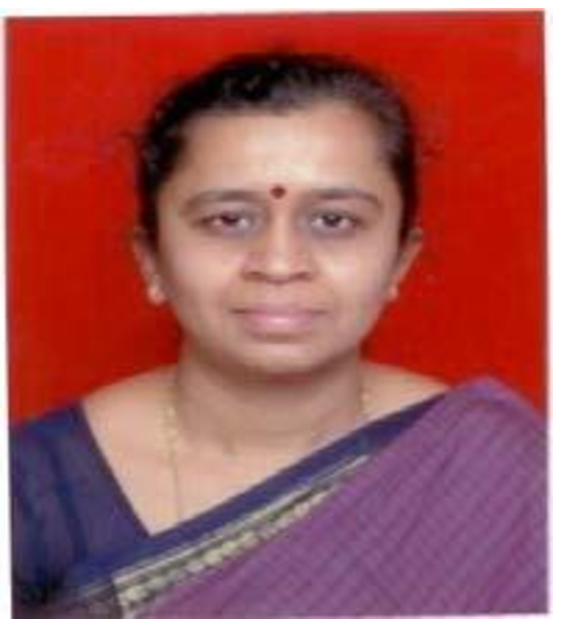}}
\begin{minipage}[b][1in][c]{1.8in}
{\centering{\bf {D N Sujatha}} is an Professor and Head in the Department of Computer Applications at B M S College of Engineering, Bangalore, India. She obtai-ned her B.Sc and MCA degrees from Mysore Univer-}\\
\end{minipage}
sity, Mysore. She obtained her Ph.D degree in the area of Computer Networks from Bangalore University. Her research interest is in the area of Sensor Networks and Computer Networks.\\\\

\noindent{\includegraphics[width=1in,height=1.7in,clip,keepaspectratio]{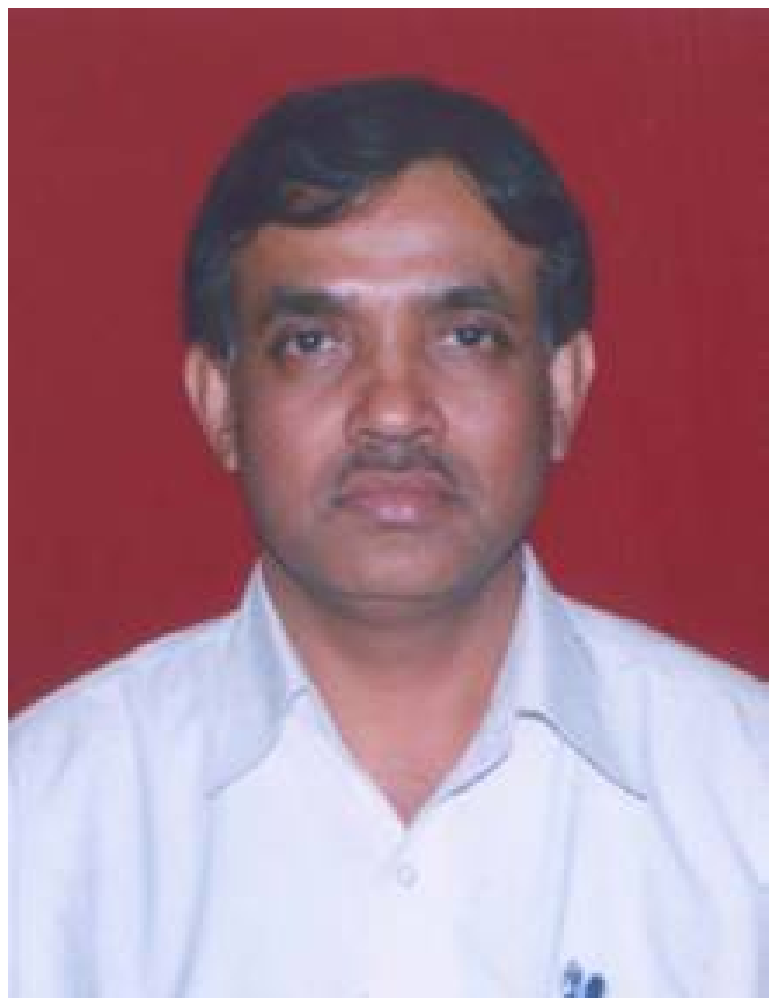}}
\begin{minipage}[b][1in][c]{1.8in}
{\centering{\bf {Venugopal K R}} is currently the Principal, University Visvesvaraya College of Engineering, Bangalore University, Bangalore. He obtained his Bachelor of Engineering from University Visvesvaraya College of Eng-}\\\\
\end{minipage}
ineering. He received his Masters degree in Computer Science and Automation from Indian Institute of Science Bangalore. He was awarded Ph.D. in Economics from Bangalore University and Ph.D. in Computer Science from Indian Institute of Technology, Madras. He has a distinguished academic career and has degrees in Electronics, Economics, Law, Business Finance, Public Relations, Communications, Industrial Relations, Computer Science and Journalism. He has authored 35 books on Computer Science and Economics, which include Petrodollar and the World Economy, C Aptitude, Mastering C, Microprocessor Programming, Mastering C++ etc. He has been serving as the Professor and Chairman, Department of Computer Science and  Engineering, University Visvesvaraya College of Engineering, Bangalore University, Bangalore. During his three decades of service at UVCE he has over 350 research papers to his credit. His research interests include Computer Networks, Parallel and Distributed Systems, Digital Signal Processing and Data Mining.\\\\
\noindent{\includegraphics[width=1in,height=1.7in,clip,keepaspectratio]{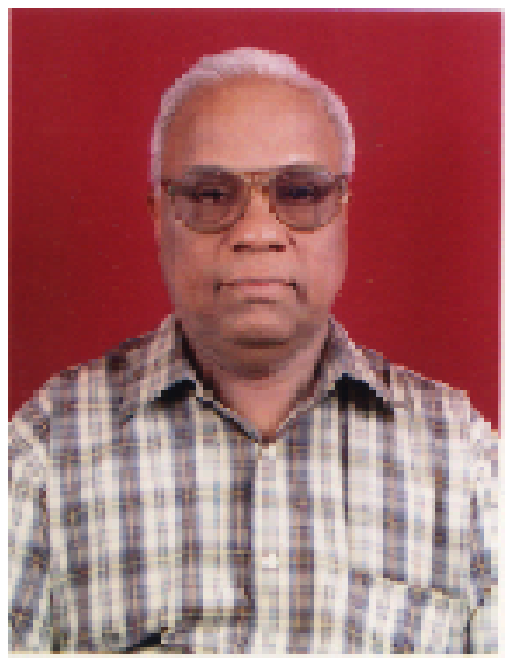}}
\begin{minipage}[b][1in][c]{1.8in}
{\centering{\bf{L M Patnaik }}is an Honorary Professor, Indian Institute of Science, Bangalore. He was the Former Vice Chancellor, Defense Institute of Advanced Technology, Pune, India. He was a Professor since 1986 with the Department of Com-}\\\\
\end{minipage}
puter Science and Automation, Indian Institute of Science, Bangalore. During the past 35 years of his service at the Institute he has over 700 research publications in refereed International Journals and referred International Conference Proceedings. He is a Fellow of all the four leading Science and Engineering Academies in India;  Fellow of the IEEE and the Academy of Science for the Developing World. He has received twenty national and international awards; notable among them is the IEEE Technical Achievement Award for his significant  contributions to High Performance Computing and Soft Computing. His areas of research interest have been parallel and distributed computing, mobile computing, CAD for VLSI circuits, Soft Computing and Computational Neuroscience.\\\\
\end{document}